\documentclass[superscriptaddress,
reprint,
 amsmath,amssymb,
 aps,
prl,
]{revtex4-2}

\usepackage{graphicx}
\usepackage{dcolumn}
\usepackage{bm}
\usepackage[hidelinks]{hyperref}
\usepackage{lineno}
\usepackage{color,soul}
\usepackage[a-1b]{pdfx}

\usepackage{easyReview}
\begin{document}

\preprint{APS/123-QED}

\title{Non-invasive 2D spatial measurements of electric fields within a plasma sheath}

\author{Mykhailo Vorobiov}
\email{Contact author: mvorobiov@wm.edu}
\author{Rob Behary}
\author{Luke Payne}
\author{Nicolas DeStefano}
\author{William Torg}
\author{Saskia Mordijck}
\affiliation{Department of Physics, The College of William \& Mary, 300 Ukrop Way, Williamsburg, VA 23185, USA}

\author{Charles T. Fancher}
\author{Neel Malvania}
\affiliation{The MITRE Corporation, McLean, Virginia 22012, USA}

\author{Edward Thomas Jr.}
\author{Saikat Chakraborty Thakur}
\affiliation{Department of Physics, Auburn University, Auburn, AL 36849, USA}

\author{Seth Aubin}
\author{Eugeniy E. Mikhailov}
\author{Irina Novikova}
\affiliation{Department of Physics, The College of William \& Mary, 300 Ukrop Way, Williamsburg, VA 23185, USA}


\date{\today}

\begin{abstract}
We introduce an all-optical quantum-enhanced diagnostic for electric fields in low-temperature plasmas. Trace amounts of rubidium vapor, added to an argon plasma, allow us to employ spectrally narrow electric field-sensitive optical resonances via the quantum optical effect of Rydberg electromagnetically induced transparency. This non-invasive technique enables measurements of electric fields in plasma with a 1~V/cm precision and a 1.5~V/cm lower detection limit. 
By collecting the fluorescence from the illuminated region of interest, we reconstruct a 2D spatial profile of the electric field magnitude with $56~\mu$m resolution.  As a proof-of-principle demonstration, we measured the changes in  the electric field within the plasma sheath surrounding a biased Langmuir probe tip. This method holds significant potential for studying sheath structures in low-temperature plasmas.

\end{abstract}

\maketitle


Electric fields (E-fields) are omnipresent in plasmas and play an important role in plasma interactions with materials~\cite{FRANKLIN2003}, in dusty plasmas~\cite{MELZER2021}, and in plasma waves, instabilities, and turbulence. Understanding plasma dynamics at the scale of the sheath and pre-sheath for plasma-material interactions requires high fidelity, non-perturbative measurements and the ability to resolve very small electric fields~\cite{HERSHOWITZ2005}. 
In the domain of high-pressure plasmas, electric-field-induced coherent anti-Stokes Raman scattering (E-CARS)~\cite{E-CARS} and electric-field-induced second harmonic generation (E-FISH)~\cite{E-FISH} techniques dominate, offering high temporal resolution of large E-fields at atmospheric pressure and above.
At lower pressures (from 10~Torr to a few mTorr), laser-induced fluorescence dip (LIF-dip)~\cite{LIF-dip, LIF-dip2, Takizawa2004} allows measurements of E-fields with a lower detection limit of $3-10$~V/cm in Ar plasmas \cite{Takizawa2002, Takizawa2004}.
Recently, dust particles suspended in low-pressure plasma using optical tweezers were employed to measure the E-field down to 4.5~V/cm~\cite{ekanayaka2025high, Ekanayaka2024}.
However, none of these techniques have been able to fully visualize the plasma sheath and pre-sheath with Debye-length precision in low-pressure plasmas. While collecting and emissive probes~\cite{HERSHOWITZ2005, HUTCHINSON2002, SHEEHAN2011} may achieve comparable spatial resolution, their own E-fields perturb the plasma environment, limiting their ability to reliably determine the E-field inside the sheath.

\begin{figure*}
    \centering
    \includegraphics[width=\linewidth]{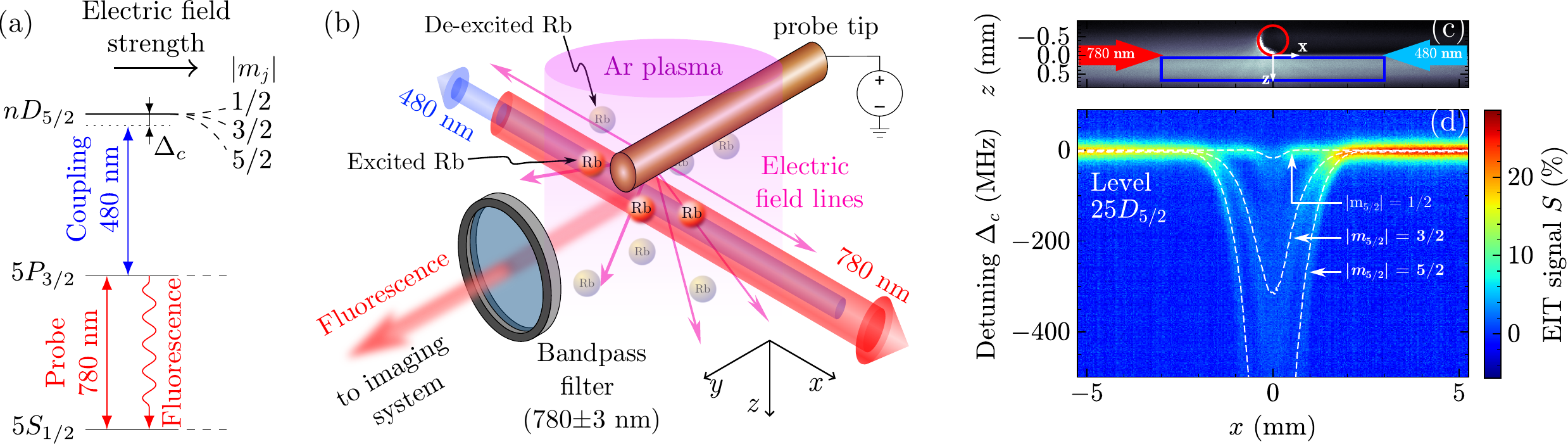}
    \caption{ 
    \textbf{(a)} Rubidium Rydberg EIT level scheme showing the Stark splitting of the $nD_{5/2}$ states versus coupling laser detuning $\Delta_c$ (defined as the difference between the $5P_{3/2}\rightarrow nD_{5/2}$ atomic transition and coupling laser frequency, so it is positive above the $nD_{5/2}$ level and negative below it). In presented experiments $n=25$.
    \textbf{(b)} Experimental schematic: counter-propagating 780~nm and 480~nm laser beams near a Langmuir probe tip in Ar plasma. Thin pink arrows indicate electrostatic field lines, thick arrows show laser propagation, and the blurred arrow indicates the detected fluorescence signal.
    \textbf{(c)} Fluorescence image showing the probe tip (red circle), coordinate axes (white arrows), and the region of interest for Fig.~\ref{fig:shrink_sheath} (blue rectangle). 
    \textbf{(d)} EIT resonances involving $25D_{5/2}$ states in Ar plasma ($n_e=1.2\times 10^{15}~\text{m}^{-3}$, $T_e = 1.7~\mathrm{eV}$, $p=14~\mathrm{mTorr}$) with the probe biased at +6.0~V. The dashed lines indicate experimentally determined positions of EIT resonances, as they split due to the Stark effect in a spatially varying $E$-field.
    \label{fig:eit-level-diagram}}
\end{figure*}

 Our method addresses these challenges by using all-optical detection and neutral ``spectator'' atoms as sensitive probes of the electrostatic plasma environment. We insert trace amounts of neutral rubidium (Rb) vapor into an argon (Ar) plasma and use a two-photon nonlinear process called electromagnetically induced transparency (EIT)~\cite{FleischhauerRevModPhys05,FinkelsteinNJP2023} to measure  minute Stark shifts of their highly excited (Rydberg) energy levels in the external E-fields \cite{Anderson2017a} as shown in Fig.~\ref{fig:eit-level-diagram}(a). Unlike the LIF-dip technique, which relies on population transfer to highly excited Rydberg states, in EIT  atoms are optically prepared in a coherent superposition of ground and Rydberg states, resulting in narrow spectral features in transmission and fluorescence. Such Rydberg EIT resonances are widely explored for detection of radio-frequency (rf) and static E-fields in an SI-traceable manner and as replacements for metal antennas and receiver systems~\cite{hollowayAPL2014,Raithel_PRA2019,Simons:18,gordonAIP2019,JauPRAppl2020,FancherIEEE2021,CoxPRL2018,prajapati2022arxiv,MeyerPRappl2021,HollowayJAP2017}, and they were employed  to measure E-fields in ion plasmas prepared from laser-cooled atom clouds~\cite{ColdPlasmaPhysRevLett.89.173004,ColdPlasmaPhysRevApplied.19.044051},
but not inside more traditional low-density ion-electron laboratory plasmas with room-temperature ions and electron temperatures, $T_e < 10$~eV. Our proof-of-principle experiment uses EIT fluorescence to map the E-field down to $\approx 56~\mu$m spatial resolution, providing a precision of 1~V/cm and a conservative lower detection limit of 1.5~V/cm, limited by the plasma-Rydberg atoms interactions. 
Successful development and deployment of this technique can enable a broad range of high-precision plasma measurements in low-temperature, low-pressure plasmas.  
In addition to sheath measurements this new technique may enable measurements of ion wakes around dust particles \cite{matthews2020dust,sametov2024influence,vermillion2024interacting}, sheath structures formed by dust clouds \cite{ivlev2006coalescence,thomas2001observations}, and potential structures formed in magnetized plasmas \cite{thomas2015quasi,hall2020dynamics, Williams2022}.

The measurement scheme is illustrated in Fig.~\ref{fig:eit-level-diagram}(b). 
Two counter-propagating laser fields illuminate the region of interest inside the plasma. 
When the sum of the laser frequencies matches the frequency difference between the ground and excited (Rydberg) states, Rb atoms are prepared in a so-called ``dark'' superposition of the ground and excited quantum states, characterized by reduced fluorescence. The electric field $E$ partially lifts the degeneracy of the $nD_{5/2}$ states, according to their $|m_j|$ value. 
In the limit of weak E-fields, the Stark shift of each sublevel depends quadratically on the amplitude of the electric field $E$ at the location of the atom:
\begin{equation}
    f_{|m_j|}^s(E) = -\frac{1}{2}\alpha_{|m_j|} E^2,
    \label{eq:stark}
\end{equation}
where $\alpha_{|m_j|}$ [in units of MHz/(V/cm)\textsuperscript{2}] denotes the polarizability of the $|m_j|$ sublevels. 
Generally, we accurately modeled Stark shifts numerically using ARC~\cite{vsibalic2017arc}.
By tracking the $|m_j|$ EIT resonances manifested as fluorescence dips, we can experimentally measure these Stark shifts and thus reconstruct the local $E$-field magnitudes~\cite{Noah2DFields,Behary2025}.



Our experimental apparatus uses low-energy electrons, accelerated by $\le36$~V voltage difference, to produce an Ar plasma via impact ionization (Ar ionization energy $\approx15.7$~eV). This method is optimal for operation at low plasma densities and temperatures \cite{Thomas_RSI2004}, while avoiding rf-fields that can impact atomic spectroscopy. In addition, the plasma can be biased by an external $E$-field.
The plasma is generated within an octagonal vacuum chamber with eight separate ports used for optical access, injection of Rb vapor, a pressure gauge, and a retractable Langmuir probe (L-probe) for plasma temperature and density measurements (see Supplementary Materials for more details). 
The same L-probe is used as the source of the E-field and sheath in the Rydberg EIT measurements.
Argon gas pressure is maintained within the range of $p=13-16$~mTorr ($\pm 15\%$), and does not cause significant collisional broadening for Rydberg EIT resonances~\cite{RybufferPhysRevApplied.23.034028,RyEITbuffer10.1063/5.0237759}. 
We use a so-called ``hot filament'' dc plasma source consisting of four 0.125~mm-diameter tungsten wires heated to the point of thermionic emission with a dc power supply with a heating current of 10~A at 2.4~V. 
The filaments are biased with respect to the grounded chamber walls, with bias voltages ranging from 0 to $-36$~V. 
This configuration produces plasma densities $n_e$ between $6\times 10^{14}~\text{m}^{-3}$ and $2\times 10^{15}~\text{m}^{-3}$, and electron temperatures $T_e$ from $2.3$ to $1.6~\text{eV}$ measured with the L-probe. 
These parameters yield the Debye screening length of $\lambda_{De}=\sqrt{\epsilon_0 T_e/e n_e} = 0.2 -0.5$~mm (where $e$ is the electron charge and $\epsilon_0$ is the vacuum permittivity), setting the scale for how far electrostatic fields penetrate plasma before quasi-neutrality is restored.
The corresponding plasma parameter $\Lambda=(4\pi\lambda_{De}^3/3) n_e$ is of order $10^5$, indicating a weakly coupled regime in which a Debye sphere of $\lambda_{De}$ radius contains a large number of electrons.
Since $\Lambda\gg1$, the plasma can be treated as ideal, with collective electrostatic interactions dominating over binary collisions between individual particles.

An oven, containing an ampule of Rb metal, is attached to one of the plasma chamber ports. The oven is heated to approximately 80~$^\circ$C for all measurements to maintain sufficient Rb vapor in the plasma chamber of approximately $5\times 10^{14}~\text{m}^{-3}$. This estimate is based on the Doppler-broadened absorption profile of the $^{85}$Rb D$_{2}$ optical transition. 
For the Ar gas pressure $p=15~\mathrm{mTorr}$ ($N_{Ar}\sim 5\times 10^{20} ~\mathrm{m^{-3}}$),  Rb atoms constitute only $\sim 1$ part in $10^6$ relative to the Ar gas and likely have a negligible impact on the plasma properties.

The optical detection consists of the counter-propagating 780 nm and 480 nm laser beams that are weakly focused to a 0.9~mm beam diameter (defined as full width at half maximum of the intensity profile). The frequency of the IR probe 780~nm laser (power 0.4~mW) is locked to the $F=3\rightarrow F'=3, 4$ cross-over resonance of the $5S_{1/2}-5P_{3/2}$ transition of $^{85}$Rb, and the blue coupling 480~nm laser (51~mW) frequency is scanned across the fine structure of the $25D$ Rydberg level. 
The calculated fine structure splitting between the $25D_{5/2}$ and $25D_{3/2}$ states of 802.215~MHz \cite{vsibalic2017arc} is used for frequency calibration, with an overall accuracy better than 1~MHz.

A simple imaging system collects the Rb fluorescence $I(x,z)$ from the laser beams passing immediately below the L-probe,  as shown by the red circle in Fig.\ref{fig:eit-level-diagram}(c). Biasing the L-probe creates a non-uniform E-field that we use to study the sheath formation.

\begin{figure}
    \centering
    \includegraphics[width=\columnwidth]{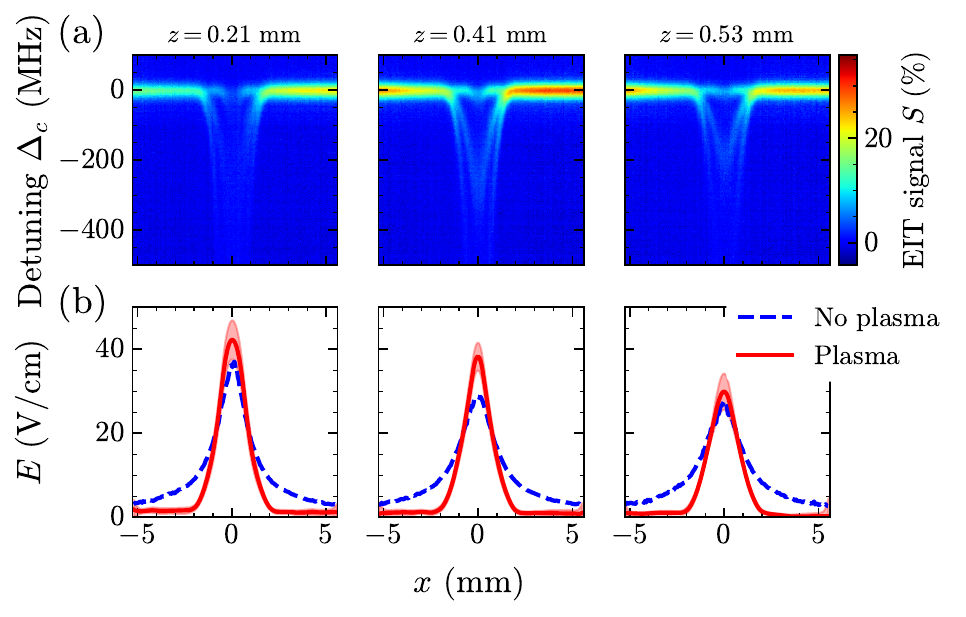}
    \caption{\textbf{(a)} Experimental maps showing evolution of the Stark-split EIT resonances at increasing vertical distances ($z$-axis) from the Langmuir probe surface ($z=0~\mathrm{mm}$) biased to +6~V in plasma. The resonances correspond to the splitting of the $25D_{5/2}$ Rydberg level. The smearing of the resonances is attributed to the electric field gradient across the cross-section of the laser beam.
    \textbf{(b)} Experimental magnitude profiles of the E-field, reconstructed from the Stark maps using Eq.~(\ref{eq:ry-eit-spectrum}) with Ar plasma (red curve; $n_e = 1.4\times 10^{15} ~\mathrm{m}^{-3}$, $T_e = 1.7~\mathrm{eV}$,  $p=14~\mathrm{mTorr}$) and without plasma (blue dashed curve). 
    The red band indicates one standard deviation uncertainty.
    }
    \label{fig:smap-z-depdence}
\end{figure}

To locate the spectral positions of the EIT resonances at each location, we scan the coupling laser frequency at a rate of 24~MHz/s (sweeping frequency 10 mHz), and record $1536\times128$~px fluorescence images with a 13.7~Hz frame rate for 683 values of the coupling laser detuning $\Delta_c$. 
Fig. \ref{fig:eit-level-diagram}(d) shows the relative change in the EIT fluorescence $S$ at 780~nm, which is the signal of interest, as a function of the position along the laser beam (horizontal $x$-axis) and the coupling laser detuning $\Delta_c$ (vertical axis). 
The EIT signal is defined as $S=1-\frac{I}{I_{bg}}$, where $I_{bg}$ is the background fluorescence signal away from the EIT resonances ($\approx +100$~MHz), the first frame in the fluorescence image sequence (top pixel row in Fig. \ref{fig:smap-z-depdence}(a)).

The EIT resonances persist even when the Ar plasma is ignited. 
Away from the L-probe, where the $E$-field is zero, we observe a single EIT peak. 
Near the probe, three resonances are clearly resolved, and their separations are largest immediately below the L-probe, where the $E$-field is the strongest. 
The resonances are noticeably broadened as a result of the $E$-field gradient across the laser beam cross-section in the $yz$-plane in Fig.~\ref{fig:eit-level-diagram}(b). 
Reducing the size of the bins along the $z$-axis in fluorescence images diminishes this broadening effect, resulting in the 2D E-field distributions as anticipated \cite{Gavrilenko2000}.

In the presence of plasma, the intrinsic EIT peak shape $L(f;\sigma)$ is altered by the electric microfields generated by the plasma ions surrounding the sensing atom. The magnitudes of these microfields are distributed according to the Holtsmark distribution \cite{Holtsmark1919, Chandrasekhar1943, griem2012spectral} defined by  Eqs.~(S1) and (S2) of Supplementary Materials (SM). 
As a result, the total electric field $E_t$ sensed the Rb atoms is a vector sum of the fluctuating microfields $E_m$ and the macroscopic field $E_e$ that we need to determine \cite{ColdPlasmaPhysRevApplied.19.044051}. Averaging over all possible microfield configurations yields a statistical distribution of the total electric field values  $P(E_t| E_e)$ for a given external electric field $E_e$ (see Eq.~(S12) in SM).

The modified fluorescence signal is a convolution of the intrinsic (homogeneous) EIT lineshape $L(f;\sigma)$ (modeled by a simple Lorentzian~\cite{MohapartaPRL2007} of full width at half maximum $\sigma$) with the electric field distribution $P(E_t|E_e)$:
\begin{equation}
\tilde{L}(f;\sigma, E_e)
=
\int_0^\infty 
P(E_t|E_e)L(f-f^s(E_t);\sigma)dE_t,
\label{eq:tilde_L_main_txt}
\end{equation}
where $f^s(E)$ is the Stark shift [defined by Eq.~(\ref{eq:stark}) for weak $E$-field and calculated numerically for arbitrary field strengths using the Alkali Rydberg Calculator (ARC) \cite{vsibalic2017arc}].
The lineshape $\tilde L$ depends on the intrinsic linewidth of the EIT $\sigma$ (that includes homogeneous broadening by the plasma electrons), the macroscopic field $E_e$, and, implicitly, on the Holtsmark field parameter $E_0 = (4\pi n_i/15)^{2/3}(e/4\pi\epsilon_0)$ estimated from the L-probe measurements of the ion density $n_i$ ($E_0$ is fixed for a given reconstruction).

To reconstruct the magnitude of the local macroscopic electric field $E_e$, we model each individual $|m_j|$ Stark component of the experimental spectra using the lineshape $\tilde L_i$ from Eq.~(\ref{eq:tilde_L_main_txt}).
Here the subscript $i$ indicates that a particular $|m_j|$ sublevel undergoes a different Stark shift $f_i^s(E)$ from the rest of the observed Stark manifold due to different polarizability $\alpha_i$.
We can then extract the value of the electric field $E_e$ by fitting the experimental EIT signal $S$ with the sum of all Stark components: 
\begin{equation}
    S(f; \sigma, E_e) = A
    \sum_{i\in\{|m_j|\}} a_i \tilde{L}_i(f; \sigma, E_e),
    \label{eq:ry-eit-spectrum}
\end{equation}
where $a_i$ represents the relative amplitudes for each of the $|m_j|$ resonances. These $a_i$ fitting parameters are normalized so that $\sum_i a_i=1$, and $A$ is an overall scaling factor. We apply the model $S$ to the entire Stark map, as illustrated in Fig.~\ref{fig:eit-level-diagram}(d). In this map, the $E_e$ field changes along the $x$ axis, while parameters $\sigma$ and $a_i$ are estimated globally for the entire map, meaning they are assumed independent of $x$.

Examples of the E-field reconstructions at three different vertical locations below the L-probe ($x=0$) are shown in Fig.~\ref{fig:smap-z-depdence}(a). 
Each fluorescence spectrum is integrated over the row of $56~\mu\mathrm{m}$ bins along $z$-axis. 
As expected, the E-field is stronger closer to the probe surface (for smaller values of $z$), resulting in more pronounced resonance shifts in the Stark maps. 
By repeating this analysis for each point across the laser beams, as shown in Fig.~\ref{fig:smap-z-depdence}(b), we can construct a 2D map of the electric field.

To illustrate the effect of the plasma on the electric field, we set the probe bias to $+6.0$~V. This value is selected to generate an E-field strong enough to clearly observe the screening effect while avoiding regions where the Stark levels become overly sensitive or mixed and would result in a poor signal.
We then gradually increase the plasma density by lowering the bias voltage of the filaments from the breakdown voltage of $-13.0$~V to $-30.0$~V at an Ar gas pressure of 14~mTorr.

\begin{figure}[h]
    \centering
    \includegraphics[width=0.9\columnwidth]{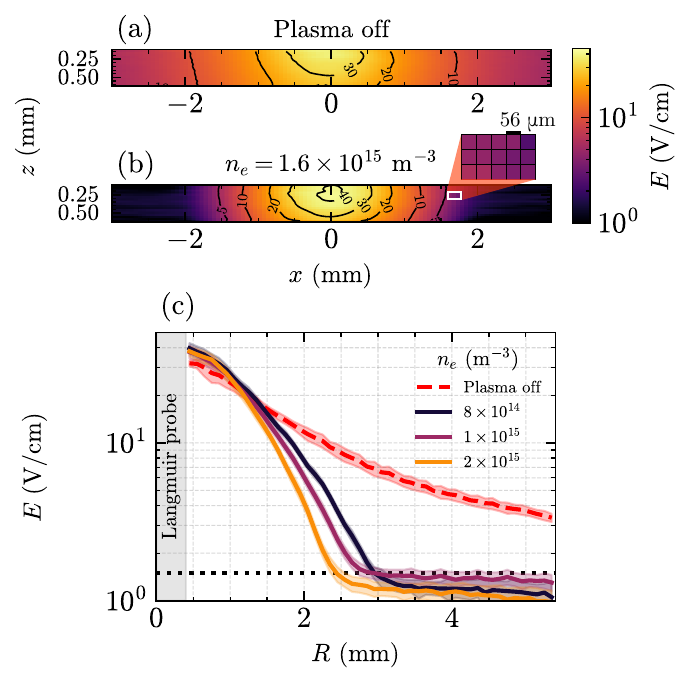}
    \caption{
    (a) 2D reconstruction of the electric field magnitude around a Langmuir probe tip biased at +6.0 V in vacuum. (b) The corresponding field distribution in the presence of an Ar plasma ($n_e=1.6\times 10^{15}~\mathrm{m}^{-3},~T_e = 2~\mathrm{eV},~p=14~\mathrm{mTorr}$). 
    The inset demonstrates a zoomed-in portion of the reconstruction (pixel size $56\times 56~\mathrm{\mu m}^2$).
    \textbf{(c)} Radial E-field profiles in the sheath of a Langmuir probe (+6.0 V bias), showing increasing screening at higher $n_e$. Profiles are derived via the angular averaging of 2D reconstructions. Shaded area: L-probe wire; dotted lines: detection limit. 
    The uncertainty bands represent the standard deviation of the radially averaged data from panels (a) and (b).
    }
    \label{fig:shrink_sheath}
\end{figure}

Figures~\ref{fig:shrink_sheath}(a,b) show examples of the 2D electric field distribution around the L-probe below and above the plasma threshold. The electric field without plasma extends further from the probe. Plasma ignition curtails the wings of the radial distribution, reducing the non-zero electric field region  to approximately 2~mm around the probe wire tip.
Figure~\ref{fig:shrink_sheath}(c) shows the evolution in the radial electric field distribution around  the L-probe wire tip surface as plasma density $n_e$ increases. The continuous reduction of the sheath electric field range clearly demonstrates the screening enhancement of the electric field with growing $n_e$. 
The observed screening enhancement is consistent with qualitative expectations. 
A quantitative theoretical validation will require detailed modeling of the L-probe and plasma chamber electrode geometry, and thus presents an important direction for future work.

The precision of the E-field reconstruction varies depending on the strength and inhomogeneity of the electric field. The technical detection limit for small fields is approximately 1.5~V/cm, limited by our ability to reliably resolve small Stark shifts. This limit makes it problematic to reliably study pre-sheath formation in this particular experiment.
For E-fields up to 15 V/cm, the uncertainties are below or around 1 V/cm thanks to the maximum coupling laser intensity  and, correspondingly, the highest EIT resonance contrast in that region ($z\approx 0.25$~mm in Fig. \ref{fig:shrink_sheath}). 
Near the L-probe surface ($z=0$~mm) the uncertainty increases again to about 3 V/cm as the signal-to-noise ratio drops.
This drop is attributed to the stronger E-field gradients near the L-probe surface and the Gaussian laser beam intensity profile, as shown by the distortions in Fig.~\ref{fig:shrink_sheath}(a,b), and to  some peaks shifting outside of the coupling laser sweep range. 
For each horizontal Stark map, the relative component amplitudes $a_i$ in Eq.~(\ref{eq:ry-eit-spectrum})
and intrinsic linewidth $\sigma$ are global parameters shared by all $x$ positions, whereas $E_e(x)$ varies independently with position as shown in Fig.~\ref{fig:eit-level-diagram}.

However, recent experiments~\cite{BeharyPRA2026} predict that the relative strength of the Stark EIT peaks depends on the orientation of the laser polarization relative to the E-field direction in each location. Thus, the lack of amplitude fitting flexibility increases the reconstruction uncertainties, especially in the regions where E-field vector orientation changes rapidly. 
These uncertainties, technical in nature, can be addressed by employing a higher power coupling beam expanded into a sheet, and by further improving the fitting model. 
In recent measurements, we demonstrated $\sim0.02$~V/cm electric field sensitivity in vacuum using the $n=58$ Rydberg level~\cite{Behary2025}. It should be possible to improve our sensitivity in plasma, enabling accurate pre-sheath measurements, although electric field fluctuations and gradients may reduce the signal-to-noise ratio. 
This issue, which requires further investigation in the context of EIT resonances, has been partially addressed recently in rf plasma~\cite{Dash2025}.

In conclusion, we demonstrate a powerful method for the electric field diagnostics inside low-temperature plasmas. By inserting trace amounts of neutral alkali metal (in our case, Rb), we can take advantage of the optical atom-based quantum sensing toolkit to image the spatial distribution of the electric field in plasma with high spatial resolution and precision. 
Specifically, we rely on the high electrostatic sensitivity of highly excited Rydberg ($n>20$) states and narrow EIT resonances to accurately map local Stark shifts. The spatial resolution of the 2D electric field reconstruction for the current experiment is around $56~\mu$m and is determined by the resolution of the imaging system for fluorescence detection and the signal-to-noise ratio of the recorded images.
We demonstrate this method by accurately monitoring electric-field variations around a biased Langmuir probe in a low-collisionality plasma, observing the expected increase in field screening with plasma density.

Future work will explore the range of achievable sensitivity of the method. In vacuum we are able to select our sensitivity from a few V/cm for $n=25$ to a few tens of mV/cm for $n\ge 45$. However, in plasmas Stark broadening from plasma microfields and Rydberg atom ionization broadens EIT resonances and limit our sensitivity~\cite{Gigosos2014}. Although higher $n$ states offer stronger field coupling, they are also more susceptible to these broadening mechanisms.
We will also explore the range of plasma conditions in which our method is applicable. For example, since collisions destroy Rydberg coherence and broaden EIT resonances~\cite{RybufferPhysRevApplied.23.034028,RyEITbuffer10.1063/5.0237759}, we expect that our approach is particularly suitable for low-pressure plasmas, but the upper limit for the gas pressure is to be established. Similar approaches have been implemented in photo-induced and inductively-coupled plasmas~\cite{ColdPlasmaPhysRevLett.89.173004,Dash2025}, suggesting that Rydberg EIT is applicable for different plasma excitation method. Finally, it will be interesting to investigate the prospects of simultaneous measurements of electric and magnetic field in magnetized plasmas, combining the tools from Rydberg electrometry and magnetometry.

This work is supported by the US Department of Energy grant SC-0024621. ET and SCT are supported by US
Department of Energy Grant SC-0019176.
Approved for Public Release by The MITRE Corporation; Distribution Unlimited. Public Release Case Number 26-0274.

\bibliography{references,bibliographyIN}

\section{Supplementary materials}
\subsection{Electric field reconstruction procedure}


\emph{Contribution of plasma-induced electric field fluctuations}: 

While the time-averaged macroscopic electric field in a neutral bulk plasma is zero, localized microscopic electric field fluctuations (plasma microfields) have a statistical distribution that smears discrete energy levels via the Stark effect. 
Because ions are heavy and move slowly, their spatial configurations remain quasi-static on the timescale of Rydberg EIT formation, yielding a microfield distribution described by the Holtsmark function. 
In contrast, the rapid motion of free electrons efficiently averages out their low-frequency spatial contributions; 
hence, their impact on the quasi-static Holtsmark microfield is negligible, and they instead manifest as a distinct homogeneous collisional broadening component \cite{griem2012spectral}.

As a result, each Rb atom experiences the total electric field $E_t$ consisting of two distinct components: an external (macroscopic) electric field $E_e$ and a microfield $E_m$ produced by the random configurations of nearby ions. In our experimental conditions the mean ion-ion distance $r_0=(3/4\pi n_{ion})^{1/3}$ (where we assume that the ion density $ n_{ion}\approx n_e$), substantially smaller than the Debye length $\lambda_{De}$ of $0.3 -0.5~\mathrm{mm}$ ($r_0/\lambda_{De}\approx 0.02 - 0.04$).
Consequently, screening is weak over the nearest-neighbor distances that dominate the high-field portion of the microfield distribution.
~\cite{Demura2010, Broyles1955}. Under these conditions the probability distribution of the microscopic field $E_m$ alone is described by the Holtsmark distribution: 
\begin{equation}
W(E_m) = \frac{1}{E_0} H(E_m/E_0),
\label{eq:W-holtsmark}
\end{equation}
where 
\begin{equation}
H(\beta)
=
\frac{2\beta}{\pi}
\int_0^\infty dx ~x\sin(\beta x)\exp(-x^{3/2}).
\label{eq:H-holtsmark}
\end{equation}
Here $\beta=E_m/E_0$ and $E_0$ is the Holtsmark field parameter related to the ion density $n_{ion}$ as
\begin{equation}
E_0 = \left(\frac{4\pi}{15}\right)^{2/3}\frac{e}{4\pi\epsilon_0}n_{ion}^{2/3},
\label{eq:Holtsmark-field}
\end{equation}
and $W(E_m)$ is normalized such that
$\int_0^\infty W(E_m)dE_m=1$.
For the present reconstruction we use the experimentally measured ion density estimated from the Langmuir probe diagnostics data, and it is fixed for the entire 2D reconstruction.
We use the isotropic Holtsmark distribution as a first-order model. 
Its accuracy in the flowing, spatially nonuniform sheath is not established and contributes a systematic uncertainty that is not included in the estimates.

Since in the experiment we measure the Stark shift averaged over many Rb atoms, we need to perform an ensemble average of the fluctuations of the microfield among all possible ion configurations.
Assuming that the ions distributed uniformly and using the Holtsmark distribution for the absolute values of the microfield $E_m=|\vec{E}_m|$, we can write the following expression for the probability of a specific value of $\vec{E}_m$:
\begin{equation}
    p(\vec{E}_m)d^3E_m
    =
    \frac{W(E_m)}{4\pi E_m^2}d^3E_m
    =
    \frac{W(E_m)}{4\pi}dE_m d\Omega ,
\end{equation}
where $d\Omega$ is the solid-angle element.

For a fixed macroscopic external field $\vec{E}_e$, the magnitude of the total field,
\begin{equation}
    E_t = |\vec{E}_e+\vec{E}_m|,
\end{equation}
depends on $E_e$, $E_m$, and the angle $\theta$ between $\vec{E}_e$ and $\vec{E}_m$ as
\begin{equation}
    E_t = g(E_m,\cos\theta;E_e)
    =
    \sqrt{E_e^2+E_m^2+2E_eE_m\cos\theta}.
\end{equation}

The probability density for observing a total field magnitude $E_t$ at a given external field $E_e$ is obtained by averaging over all microfield magnitudes and orientations:
\begin{eqnarray}
   && p(E_t|E_e)
    =\\&&
    \int_0^\infty dE_m \frac{W(E_m)}{4\pi}
    \int d\Omega\,
    \delta\left[
    E_t-g(E_m,\cos\theta;E_e)
    \right] = \nonumber \\&&
     \int_0^\infty dE_m \frac{W(E_m)}{2}
    \int_{-1}^{1} d\xi\,
    \delta\left[
    E_t-g(E_m,\xi;E_e)
    \right],\nonumber
\end{eqnarray}
where $\xi=\cos\theta$.

For fixed $E_e$ and $E_m$, the argument of the delta function has a root at
\begin{equation}
    \xi^\ast
    =
    \frac{E_t^2-E_e^2-E_m^2}{2E_eE_m}.
\end{equation}
Using
\begin{equation}
    \delta\left[E_t-g(E_m,\xi;E_e)\right]
    =
    \frac{\delta(\xi-\xi^\ast)}
    {\left|\partial g/\partial \xi\right|_{\xi=\xi^\ast}},
\end{equation}
and
\begin{equation}
    \left.\frac{\partial g}{\partial \xi}\right|_{\xi=\xi^\ast}
    =
    \frac{E_eE_m}{E_t},
\end{equation}
we obtain
\begin{equation}
    \int_{-1}^{1} d\xi\,
    \delta\left[
    E_t-g(E_m,\xi;E_e)
    \right]
    =
    \frac{E_t}{E_eE_m},
\end{equation}
using $|E_t-E_e| \leq E_m \leq E_t+E_e $.

The resulting probability density of the total electric-field magnitude is
\begin{equation}
    p(E_t|E_e)
    =
    \frac{E_t}{2E_e}
    \int_{|E_t-E_e|}^{E_t+E_e}
    \frac{W(E_m)}{E_m}\,dE_m .
    \label{eq:P_Et_Ee}
\end{equation}
Note that in the limiting case of zero macroscopic field, this expression reduces to
\begin{equation}
    p(E_t|0)=W(E_t),
\end{equation}
as expected, since the total field is then determined solely by the ion microfield.

\emph{Effect of plasma-induced electric field distribution on the EIT lineshape}:
We assume the base lineshape of the EIT resonance $L_\sigma$ broadened by plasma electrons to be a simple Lorentzian
\begin{equation}
    L(f-f^s(E);\sigma) = \frac{\sigma/2\pi}{(f-f^s(E))^2 + (\sigma/2)^2},
    \label{eq:Lorentzian}
\end{equation}
where $f^{s}(E)$ is the Stark shift produced by a uniform electric field $E$, and $\sigma$ is the full width at half maximum (FWHM) of the EIT peak when quasi-static ion fields are absent, but it includes additional broadening caused by electrons \cite{griem2012spectral}. 
If necessary, the more accurate base lineshape can be calculated using Lindblad equation \cite{Dash2025} at the expense of increased calculational complexity. 

In the presence of ionic microfields, the local variations of the total electric field, sensed by Rb atoms, introduces additional inhomogeneous broadening, and the resulting EIT lineshape $\tilde{L}(f;\sigma, E_e)$ is modeled as a convolution of the distribution $P(E_t| E_e)$ and the base EIT lineshape: 
\begin{equation}
\tilde{L}(f;\sigma, E_e)
=
\int_0^\infty 
P(E_t|E_e)L(f-f^s(E_t);\sigma)dE_t.
\label{eq:tilde_L}
\end{equation}
Note that $\tilde L$ implicitly depends on the Holtsmark field parameter $E_0$ defined by Eq.~(\ref{eq:Holtsmark-field}) through $P(E_t|E_e)$ distribution, and the dependence $f^s(E)$ is defined by the specific atomic Rydberg states.

The complete spectrum is calculated as the sum of individual lineshapes (\ref{eq:tilde_L}) for all accessible Rydberg levels:
\begin{equation}
    S(f; E_e) = \sum_i a_i \tilde{L}_i(f; \sigma, E_e),
    \label{eq:S-spectrum-model}
\end{equation}
where the relative intensities of the EIT resonances $a_i$ are the global fitting parameters  for the entire Stark map.

\emph{Contribution of spatial gradients}:
In addition to the broadening of the EIT resonances caused by plasma electric-field fluctuations, the measured spectra exhibit further broadening due to spatial variations of the electric field within a single camera pixel. 
When the electric-field gradient is sufficiently large, the Stark shift varies across the pixel area, resulting in an effective averaging of resonances with different center frequencies.

To account for this effect, we model the observed spectrum by convolving the lineshape $\tilde{L}(f;\sigma,E)$ from Eq.~(\ref{eq:tilde_L}) with a Gaussian kernel $G(f,w)=\exp{(-f^2/2w^2)}/\sqrt{2\pi w^2}$ where the width $w$ represents the frequency spread induced by unresolved electric-field variations within a pixel. 
The resulting observed lineshape is defined as
\begin{equation}
    \tilde L^{obs}(f; \sigma, E)
    =
    \int df_0
    \tilde L(f_0; \sigma, E)G(f-f_0, w).
    \label{eq:S-gradient}
\end{equation}
The Gaussian width is related to the local electric-field variation through $w = |df^s(E)/dE| \delta E$
where $df^s/dE$ is the theoretical Stark-shift sensitivity and $\delta E$ is the effective electric-field variation across a pixel, parameterized as $\delta E = \sqrt{ (\Delta z\cdot dE/dx)^2 + (\mathcal{G}_\perp \cdot E)^2}$ where $\Delta z$ is the pixel size, $dE/dx$ is the spatial derivative of the reconstructed electric-field profile, and $\mathcal{G}_\perp$ is a global fitting parameter that accounts for electric-field gradients in directions not explicitly included in the one-dimensional reconstruction.
The first term therefore describes broadening due to the resolved field gradient along the measurement axis, while the second term captures unresolved transverse gradients

\begin{figure}
    \centering
    \includegraphics[width=1\linewidth]{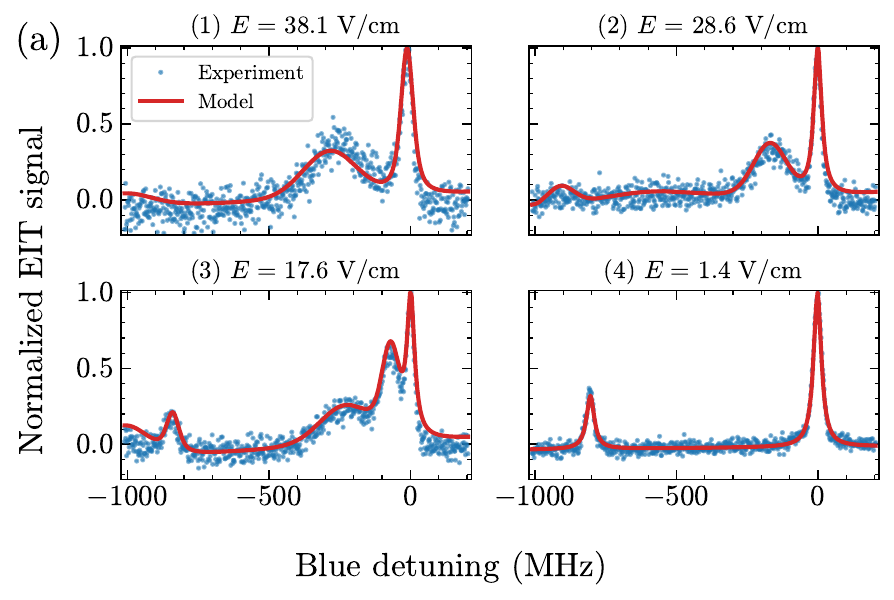}
    \includegraphics[width=0.9\linewidth]{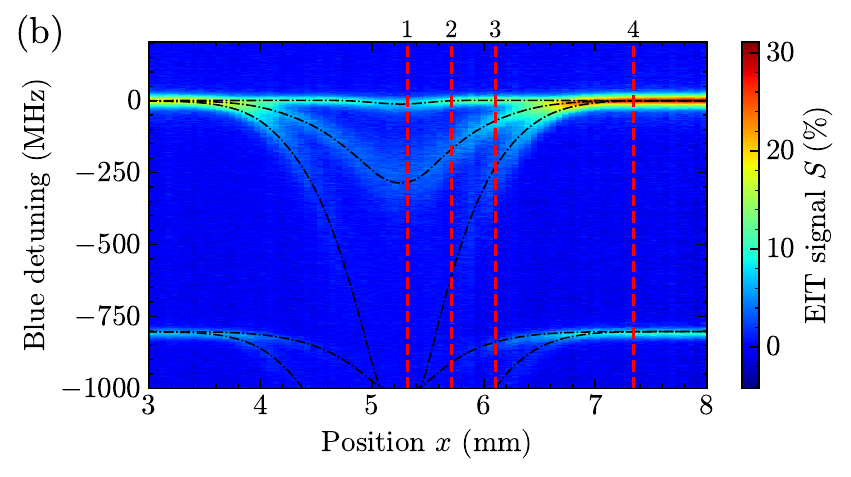}
    \caption{
    \textbf{(a)} Representative spatially resolved EIT spectra extracted from the Stark map at the numbered positions indicated in the lower panel. The red curves show model EIT spectra calculated using Eqs.~(\ref{eq:S-spectrum-model}) and (\ref{eq:S-gradient}).
    \textbf{(b)} Position-dependent EIT Stark map measured in an Ar plasma with electron density $n_e = 1.4\times10^{15}~\mathrm{m}^{-3}$, electron temperature $T_e = 1.7~\mathrm{eV}$, and pressure $p = 14~\mathrm{mTorr}$. The data were acquired at a vertical distance of $z = 0.41$~mm from the Langmuir probe surface. The white dash-dotted curves indicate the extracted Stark-shift trajectories of the EIT resonance as a function of position. The numbered labels at the top of the map identify the spatial locations from which the spectra shown in panel (a) were sampled.
    }
    \label{fig:SM-repr_spectra_and_fitted_smap}
\end{figure}

Although Eq.~(\ref{eq:S-gradient}) is an approximate description of the underlying spatial averaging process, it reproduces the overall scaling of gradient-induced broadening. 
As shown in Fig.~\ref{fig:SM-repr_spectra_and_fitted_smap}(a), the model provides a reasonable description of the measured spectra.
The main discrepancy arises because the measurement is not directly sensitive to the full electric-field vector. In particular, the EIT resonance amplitudes depend on the orientation of the laser polarizations with respect to the local electric-field vector \cite{BeharyPRA2026}, an effect that is not captured by the present model.
These polarization-dependent effects are not included in the present model.

Taking these limitations into account, we conservatively estimate the statistical fit uncertainty as $\leq 1~\mathrm{V/cm}$. Systematic uncertainty associated with the lineshape and polarization model has not yet been fully quantified.
Near the maximum of the reconstructed field, however, the uncertainty increases to approximately 3~V/cm. 
In this regime, some Stark-shifted resonances extend beyond the scanned coupling-laser detuning range, reducing the constraints imposed by the data on the fit parameters. 
An example of this behavior can be seen in Fig.~\ref{fig:SM-repr_spectra_and_fitted_smap}(a), spectrum (1).

\emph{Comparison with the experimental data}:
The fitting uses the Levenberg-Marquardt minimizer algorithm from the ``lmfit'' package \cite{LMFIT_reference} on a horizontal slice of the fluorescence images, as shown in Fig.~\ref{fig:SM-repr_spectra_and_fitted_smap}(b) (b) of SM and Fig 1(c) of the main text. At each position $z$ from the L-probe's surface, the row of pixels of the fluorescence image along the $x$ coordinate is expanded into the Stark map shown in Fig.~\ref{fig:SM-repr_spectra_and_fitted_smap}(d). 
The model (\ref{eq:S-spectrum-model}), adjusted by the technical gradient (\ref{eq:S-gradient}), is then fitted to the entire Stark map, not to individual spectra at each $x$. 
In short, during each minimizer iteration, a new Stark map is generated based on the model (\ref{eq:S-spectrum-model}) which is two-dimensional array of EIT spectra for each $x$-position.
The algorithm minimizes the difference $\Delta S_{ij}$ between the data $S_{ij}^{data}$ and the model $S_{ij}^{model}$  weighted by the uncertainties $u_{ij}$ with respect to the Frobenius norm
$||\Delta S_{ij}/u_{ij}||_F^2 = \sum_{ij} (\Delta S_{ij}/u_{ij})^2$,
where the indices $i$ and $j$ represent EIT signal $S$ samples along the frequency detuning $f$ and spatial $x$ axes, respectively.
This allows for a global estimation of the relative intensities $a_i$ of the EIT resonances defined in Eq.~(\ref{eq:S-spectrum-model}) and the intrinsic EIT resonance width $\sigma$. The macroscopic electric field $E_e$ is varied along the $x$ coordinate, and its one-dimensional distribution is extracted.
The uncertainties $u_{ij}$ were estimated from the experimental noise by first detrending each spectrum with a median filter ($10~\mathrm{px}$ window) and subsequently calculating the moving standard deviation of the residuals ($10~\mathrm{px}$ window).
The windows were chosen to be larger than the pixel-scale noise correlations but smaller than the characteristic spatial and spectral variation scales of the EIT features.

\subsection{Radial averaging under cylindrical symmetry}

\begin{figure}[h]
    \includegraphics[width=0.6\linewidth]{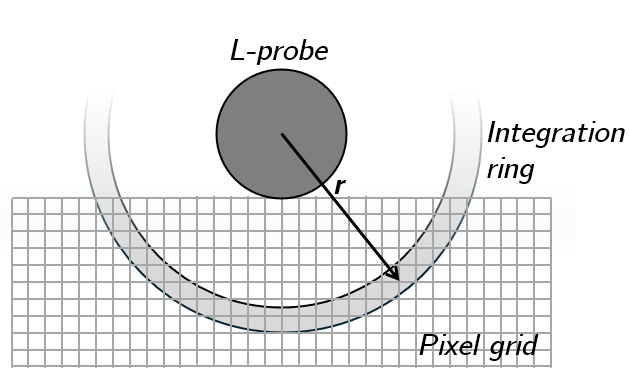}
    \caption{Illustration of radial averaging of the electric field}
    \label{fig:R-unfolding}
\end{figure}

The angular averaging procedure used to produce Fig.~3(b) (main text) is auxiliary and helps to unfold the spatially restricted reconstruction within the finite laser beam into a full radial E-field distribution around the Langmuir probe.
It is done numerically 
relying on cylindrical symmetry of the Langmuir probe used in this experiment. 
The algorithm is illustrated in Fig.~\ref{fig:R-unfolding} and is as follows:
(i)~calculate how far away each pixel from the center of the probe tip circular area.
(ii)~create a series of target concentric rings (``bins'') spreading out from the center point up to a maximal distance.
(iii)~calculate how many pixels landed inside each concentric ring and average pixels within the ring.
Pixels that do not contain reconstructed E-field samples are not included in the averaging.
This approach allowed us extract the radial 1D E-field reconstruction for up to $\sim 5~\textrm{mm}$ which is half of the total reconstructed distance along horizontal $x$ axis in the original fluorescence image (width: 10.8~mm).

\subsection{Plasma setup}

\begin{figure}[h]
    \centering
    \includegraphics[width=\linewidth]{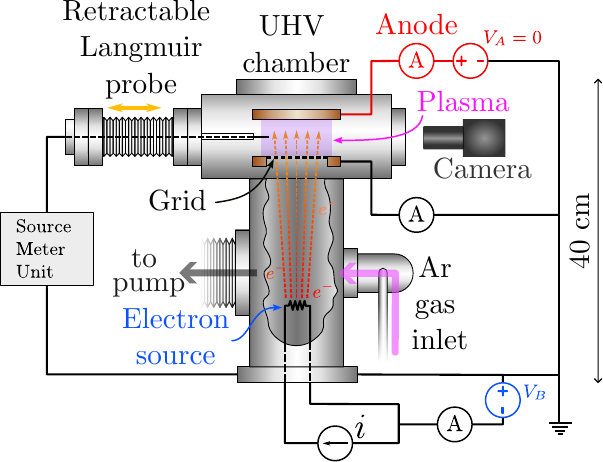} 
    \includegraphics[width=1\linewidth]{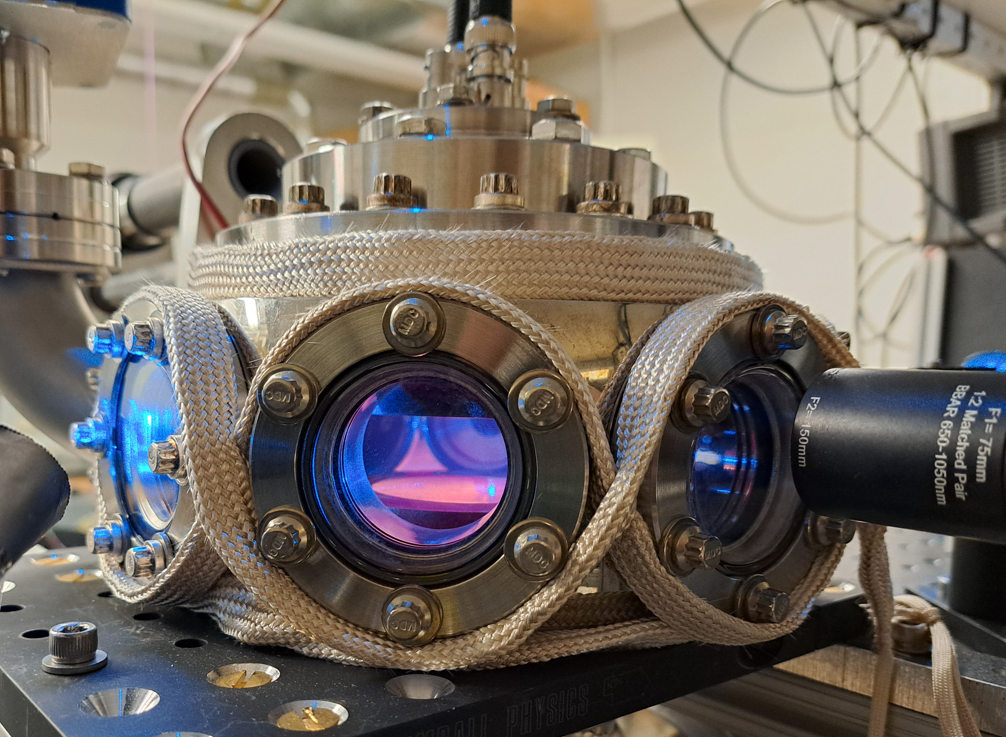}
    \caption{
    \textbf{(a)} Schematic of the complete UHV setup. Electrons emitted from the thermionic tungsten filaments are accelerated toward the upper chamber region by applying a negative bias voltage $V_B$ to the electron source relative to the grounded chamber walls. An independent anode voltage supply, $V_A$, is shown for completeness but was not used in the experiments reported here.
    \textbf{(b)} Enlarged view of the upper chamber region where plasma spectroscopy is performed. The measurement volume is located between the copper anode and tungsten grid electrodes, with a Langmuir probe positioned between them for plasma diagnostics and electric-field perturbation studies. Optical excitation and fluorescence collection are performed through opposing viewports.
    }
    \label{fig:setup_schematics}
\end{figure}
The thermionic plasma generation apparatus consists of an ultra-high-vacuum (UHV) stainless-steel chamber, shown in Fig.~\ref{fig:setup_schematics}. 
The upper section of the system shown in Fig.~\ref{fig:setup_schematics}(b) is an octagonal chamber equipped with eight viewports that provide optical access. 
Three of these ports are used for a vacuum gauge, a retractable Langmuir probe assembly, and a 1.33-inch vacuum T-connector sealed at both ends and used as a Rb oven containing a 1~g ampoule of metallic rubidium.

Ultra-high-purity argon gas (99.9999\%) is introduced into the chamber through a leak valve. The chamber pressure is monitored using an MKS390 vacuum gauge.

The electron source consists of four tungsten filaments clamped between copper holders mounted on a ceramic base. 
The filaments are heated by a current source to produce thermionic emission, with an operating current of 10~A. 
The emitted electrons are accelerated toward the upper section of the chamber, shown separately in Fig.~\ref{fig:setup_schematics}(b), where the spectroscopic measurements are performed.

Electromagnetically induced transparency (EIT) measurements are carried out in the region between two electrodes separated by 24~mm. 
The upper (``anode'') electrode is made of copper, while the lower electrode is a tungsten mesh grid. 
A Langmuir probe is positioned midway between the electrodes to characterize the plasma parameters and to generate a localized inhomogeneous electric field for the experiment. 
The probe tip is fabricated from copper wire and has a length of 14.4~mm and a diameter of 0.8~mm.
The Langmuir probe bias and current measurements were controlled using a source meter unit (Aim-TTi SMU4201).

Optical excitation is provided by two TOPTICA DL pro diode lasers. 
The resulting fluorescence is detected using an industrial infrared camera (FLIR BFS-PGE-63S4M) with $2\times 2$ hardware binning. 
The fluorescence signal is collected through a matched pair of achromatic doublets (MAP1075150-B, Thorlabs).
The imaging system provided a field of view of $10.8\times7.2~\mathrm{mm}$.


\end{document}